\documentclass[10pt,journal,final]{IEEEtran}
\usepackage{mathrsfs}
\usepackage{bbding}
\usepackage{graphicx}
\usepackage{float}
\usepackage{booktabs}
\usepackage[hidelinks]{hyperref}
\usepackage{cite}

\usepackage{enumitem}
\usepackage{color}
\usepackage{ulem}
\usepackage{psfrag}
\usepackage{subfigure}
\usepackage{amssymb}
\usepackage{amsthm}
\usepackage{setspace}
\usepackage{epsfig}
\usepackage{pifont}
\usepackage{amsmath}
\usepackage{array}
\newcommand{\PreserveBackslash}[1]{\let\temp=\\#1\let\\=\temp}
\newcolumntype{C}[1]{>{\PreserveBackslash\centering}p{#1}}
\newcolumntype{R}[1]{>{\PreserveBackslash\raggedleft}p{#1}}
\newcolumntype{L}[1]{>{\PreserveBackslash\raggedright}p{#1}}
\usepackage{multicol}
\usepackage{multirow}
\usepackage{diagbox}
\usepackage{pifont}
\usepackage{indentfirst}
\usepackage{amsfonts}
\usepackage{algorithm}
\usepackage{algorithmicx}
\usepackage{algpseudocode}
\floatname{algorithm}{Procedure}
\usepackage{fancyhdr}
\usepackage{amscd}
\usepackage[cmyk]{xcolor}
\usepackage{bm}
\usepackage{fancyhdr}
\setlist{itemsep=0pt,parsep=0pt}

\newtheorem{proposition}{Proposition}

\allowdisplaybreaks[4]

\IEEEoverridecommandlockouts

\usepackage{caption}
\begin{document}
	\title{\huge GPASS: Deep Learning for Beamforming in\\ Pinching-Antenna Systems (PASS)}
	
	\author{
		\thanks{Jia Guo is with the School of Electronic Engineering and Computer
			Science, Queen Mary University of London, London E1 4NS, U.K. (e-mail:
			jia.guo@qmul.ac.uk).
			
			Yuanwei Liu is with the Department of Electrical and Electronic Engineering,
			The University of Hong Kong, Hong Kong (e-mail: yuanwei@hku.hk).
			
			Arumugam Nallanathan is with the School of Electronic Engineering and Computer
			Science, Queen Mary University of London, London E1 4NS, U.K. (e-mail:
			a.nallanathan@qmul.ac.uk).
		}
		\IEEEauthorblockN{Jia Guo, Yuanwei Liu, \emph{Fellow, IEEE}, and Arumugam Nallanathan, \emph{Fellow, IEEE}}
		
	}
	\maketitle
	\setcounter{page}{1}
	\thispagestyle{empty}
	
	\begin{abstract}
		A novel GPASS architecture is proposed for jointly learning pinching beamforming and transmit beamforming in pinching antenna systems (PASS). The GPASS is with a staged architecture, where the positions of pinching antennas are first learned by a sub-GNN. Then, the transmit beamforming is learned by another sub-GNN based on the antenna positions. The sub-GNNs are incorporated with the permutation property of the beamforming policy, which helps improve the learning performance. The optimal solution structure of transmit beamforming is also leveraged to simplify the mappings to be learned. Numerical results demonstrate that the proposed architecture can achieve a higher SE than a heuristic baseline method with low inference complexity.
		
	\end{abstract}

	\section{Introduction}\label{sec:intro}
	As a novel member of the flexible-antenna system family, the pinching-antenna system (PASS) offers significant potential for revolutionary advancements in wireless communications. The core concept involves applying low-cost dielectric materials at arbitrary locations on dielectric waveguides, creating stable line-of-sight (LoS) links to enhance channel conditions. Compared to existing flexible-antenna systems, such as fluid or movable antennas \cite{nearfield-survey}, PASS provide greater flexibility in large-scale adjustments to mitigate the impact of non-LoS links caused by blockages \cite{ding2024flexible}.
	
	In PASS, the optimization related to the pinching antennas is critical to improving system performance such as spectral efficiency (SE) and energy efficiency. In \cite{ouyang2025array}, it was revealed that the number of pinching antennas and the inter-antenna spacing can be optimized to maximize the array gain. In \cite{wang2024antenna}, a SE-maximization problem in the PASS was optimized, from which the antennas at desired positions can be activated. The pinching beamforming (which depends on the positions of pinching antennas) and resource allocation or transmit beamforming can also be jointly optimized. These optimization problems can be challenging, because i) the optimization of the two types of variables is coupled, ii) the relationship between the pinching antenna positions and the phases of channel coefficients is non-convex. and iii) the interference introduces additional non-convexities. In \cite{liu2025pinching}, a penalty dual decomposition-based method was proposed to optimize the transmit and pinching beamforming, aiming to minimize the transmit power in the system.

	Noticing that deep neural networks (DNNs) are good at learning unknown mappings from massive data, existing works designed DNNs to
	learn wireless policies such as beamforming and power allocation, including fully
	connected neural networks (FNNs) \cite{L2O}, convolutional
	neural networks (CNNs) and graph neural networks
	(GNNs) \cite{ENGNN,GNN_RIS_FL_TWC_2024,GJ-RGNN,LSJ_MultiDim_GNN_2022,GNN_mUE_RIS_JSAC_2021_Yu}. Among these, GNNs have demonstrated superior performance due to their ability to i) generalize to unseen graph sizes \cite{GNN_RIS_FL_TWC_2024}, ii) achieve better learning performance with fewer samples \cite{GJ-RGNN,ENGNN} and iii) scale efficiently to large systems \cite{GNN-PC-CellFree-TWC2024,GNN_mUE_RIS_JSAC_2021_Yu}. 
	It has been noticed that a key factor contributing to the effectiveness of GNNs is their ability to leverage the permutation properties inherent in wireless policies \cite{LSJ_MultiDim_GNN_2022}.
	
	Inspired by the success of GNNs in learning wireless policies, this letter proposes a novel GNN-based architecture for jointly optimizing pinching beamforming and transmit beamforming in PASS, which is referred to as GPASS. To the best of our knowledge, this is the first study to introduce deep learning for PASS. Noticing that simultaneously outputting the pinching and transmit beamforming can be challenging, the GPASS first learns the pinching beamforming, and then learns the transmit beamforming with two distinct sub-GNNs. The sub-GNNs can harness permutation properties satisfied by the mappings to be learned, thereby improving learning performance.
	Numerical results demonstrate the performance of the proposed architecture in terms of achieving high SE with low inference complexity.
	
	\emph{Notations}: $(\cdot)^T$ and $(\cdot)^{H}$ respectively denote the transpose and conjugate transpose of a matrix, $\mathbf{X}^{\star}$ denotes the optimal value of variable $\mathbf{X}$.
	
	\section{System Model and Problem Formulation}
\begin{figure}[!htb]
		\centering
		\includegraphics[width=\linewidth]{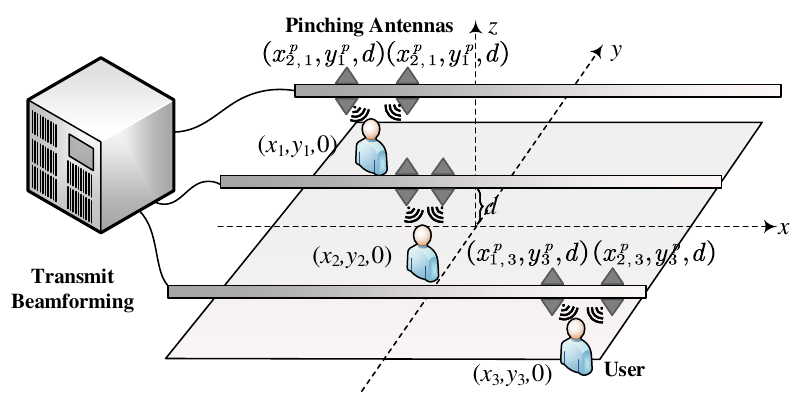}
		\caption{Illustration of PASS, $K=3, N=3, M=2$.}
		\label{fig:pinching}
		\vspace{-5mm}
	\end{figure}

	Consider a downlink system as shown in Fig. \ref{fig:pinching}, where a base station (BS) equipped with $N$ waveguides transmits to $K$ single antenna users. There are $M$ pinching antennas on each waveguide. The users are deployed in a squared area with length $D$ on the $x-y$ plane. The position of the $k$-th user is denoted as $\bm\psi_k=(x_k, y_k, 0)$. Without the loss of generality, it is assumed that the waveguides of the BS are deployed at the height of $d$, and are parallel to the $x$-axis. The length of each waveguide is the same as the side of the squared area. The position of the $m$-th pinching antenna on the $n$-th waveguide is denoted as $\bm\psi_{m,n}^p=(x_{m,n}^{p}, y_n^{p}, d)$. 
	
	For the pinching antennas on the same waveguide, the transmitted signal of each pinching antenna is a phase-shifted version of the signal transmitted at the feed-point of the waveguide. Then, the transmitted signal at the $k$-th user is denoted as,
	\begin{equation}
		\mathbf{s}_k = \underbrace{\mathbf{G}}_{\text{\parbox{1.4cm}{\centering Pinching beamforming}}}\cdot\underbrace{\mathbf{w}_k}_{\text{\parbox{1.4cm}{\centering Transmit beamforming}}}\cdot~~ x_k,
	\end{equation}
	where $x_k$ is the transmitted symbol of the $k$-th user, $\mathbf{w}_k\in{\mathbb C}^{N\times 1}$ is the transmit beamforming vector for the $k$-th user, and 
	\begin{equation}\label{eq:pinching-bf}
		\mathbf{G} = \begin{bmatrix}
			\mathbf{g}_1 & \cdots & 0\\
			\vdots & \ddots & \vdots \\
			0 & \cdots & \mathbf{g}_N
		\end{bmatrix},
	\end{equation}
	$\mathbf{g}_n=[e^{-j\frac{2\pi}{\lambda_g}\|\bm\psi_{0,n}^p-\bm\psi_{1,n}^p\|},\cdots,e^{-j\frac{2\pi}{\lambda_g}\|\bm\psi_{0,n}^p-\bm\psi_{M,n}^p\|}]^T$.
	Furthermore, $\bm\psi_{0,n}^p$ is the position of the feed-point of the $n$-th waveguide, and $\lambda_g=\lambda/n_{\rm eff}$ is the guide wavelength with $\lambda$ being the wavelength in the free space and $n_{\rm eff}$ being the effective refractive index of the dielectric waveguide. In what follows, we refer to  $\mathbf{G}$ as the \emph{pinching beamforming matrix}.
	
	The received signal at the $k$-th user is denoted as,
	\begin{equation}\label{eq:receive-signal}
		y_k = \textstyle\mathbf{h}_k^H \mathbf{s}_k + \sum_{j=1,j\neq k}^K \mathbf{h}_k^H \mathbf{s}_j + n_k,
	\end{equation}
	where $\mathbf{h}_k = [\mathbf{h}_{1k},\cdots,\mathbf{h}_{Nk}]^T$ is the channel from the pinching antennas to the $k$-th user, and 
	$\mathbf{h}_{nk}$ is the channel vector from the $n$-th pinching antenna to the $k$-th user. It can be expressed as,
	\begin{equation}\label{eq:channel-vec}
		\mathbf{h}_{nk} = \left[\frac{\sqrt{\eta}e^{-j\frac{2\pi}{\lambda}\|\bm\psi_k-\bm\psi_{1,n}^p\|}}{\|\bm\psi_k-\bm\psi_{1,n}^p\|},\cdots,\frac{\sqrt{\eta}e^{-j\frac{2\pi}{\lambda}\|\bm\psi_k-\bm\psi_{M,n}^p\|}}{\|\bm\psi_k-\bm\psi_{M,n}^p\|}\right],
	\end{equation}
	where $\eta=\frac{c}{2\pi f_c}$ is a constant with $c$, $f_c$ and $\lambda$ denoting the speed of light, the carrier frequency and the wavelength in the free space, respectively.
	
	The positions of the pinching antennas and the transmit beamforming can be optimized to maximize the SE of the system. The optimization problem can be formulated as,
	\begin{subequations}\label{eq:opt}
		\begin{align}
			\max_{\bm\Phi^p, \mathbf{W}} ~&\sum_{k=1}^K \log_2\Bigg(\!1+\frac{|\mathbf{h}_k^H\mathbf{Gw}_k|^2}{\sum_{j=1,j\neq k}^K |\mathbf{h}_k^H\mathbf{Gw}_j|^2+\sigma_0^2}\!\Bigg) \label{eq:opt-objective}\\ 
			{\rm s.t.} ~& 0\leq x_{m,n}^{p} \leq D, \forall m,n, \label{eq:opt-constraint-01}\\
			&\Delta_{m,n}^p\triangleq x_{m,n}^p-x_{m-1,n}^p\geq \Delta_{\min}, \forall m, \label{eq:opt-constraint-03}\\
			& \|\mathbf{GW}\|^2 = \|\mathbf{W}\|^2 \leq P_{\max}, \label{eq:opt-constraint-2}
		\end{align}
	\end{subequations}
	where $\bm\Phi^p=[\bm\psi_{1,1}^p,\cdots,\bm\psi_{M,N}^p]$ is the vector of positions of all the pinching antennas, $\mathbf{W}=[\mathbf{w}_1,\cdots,\mathbf{w}_K]$ is the baseband precoding matrix.
	$\sigma_0^2$ is the noise power, \eqref{eq:opt-constraint-03} restricts the minimum distance among pinching antennas to avoid mutual coupling,
	and \eqref{eq:opt-constraint-2} is the constraint that the transmit power cannot exceed a maximal power budget $P_{\max}$.
	
	To satisfy the constraint in \eqref{eq:opt-constraint-3} with the learning-based method to be proposed later, we resort to the relationship $x_{m,n}^p=x_{1,n}^p + \sum_{i=2}^m \Delta_{m,n}^p$ and transform the problem in \eqref{eq:opt} to the one optimizing $\mathbf{x}_1^p=[x_{1,1}^p,\cdots,x_{1,N}^p]^T$ and $\bm\Delta^p=[\Delta_{1,1}^p,\cdots,\Delta_{M,N}^p]$ as follows,
	\begin{subequations}\label{eq:opt-1}
		\begin{align}
			\mathsf{P}: ~\max_{\mathbf{x}_1^p, \bm\Delta^p} ~& \eqref{eq:opt-objective} \notag\\ 
			{\rm s.t.} ~& 0\leq x_{1,n}^p + \sum_{i=2}^m \Delta_{m,n}^p \leq D, \forall m,n, \label{eq:opt-constraint-1}\\
			&\Delta_{m,n}^p \geq \Delta_{\min}, \forall m, \label{eq:opt-constraint-3}\\
			& \eqref{eq:opt-constraint-2}. \notag
		\end{align}
	\end{subequations}
	
	Given a set of user positions $\bm\Phi=[\bm\psi_1,\cdots,\bm\psi_K]$, problem \textsf{P} can be solved to optimize $\mathbf{x}_1^p, \bm\Delta^p$ and $\mathbf{W}$. Denote the mapping from the user positions to the optimal variables as $\{\mathbf{x}_1^{p\star},\bm\Delta^{p\star},\mathbf{W}^\star\}=F(\bm\Phi)$, which is called \emph{beamforming policy} in the sequel.
	
	By finding the permutations that do not affect the objective function and constraints of problem \textsf{P}, it is not hard to prove that the beamforming policy satisfies the following permutation property,
	\begin{equation}\label{eq:pe-policy}
		\{\mathbf{\Pi}_{\sf A}^T\mathbf{x}_1^{p\star},\mathbf{\Omega}_{\mathsf{P}}^T\bm\Delta^{p\star},\mathbf{\Pi}_{\mathsf{A}}^T\mathbf{W}^\star\mathbf{\Pi}_{\mathsf{U}}\}=F(\mathbf{\Pi}_{\mathsf{U}}^T\bm\Phi),
	\end{equation}
	where $\mathbf{\Pi}_{\sf A}$ is an arbitrary permutation matrix that change the order of waveguides,  $\mathbf{\Omega}_{\mathsf{P}}=\mathbf{\Pi}_{\mathsf{A}}\otimes\mathsf{diag}(\mathbf{\Pi}_{\mathsf{P},1},\cdots,\mathbf{\Pi}_{\mathsf{P},N})$ is a nested permutation matrix that changes the order of waveguides by $\mathbf{\Pi}_{\mathsf{A}}$ and changes the order of pinching antennas in the $n$-th waveguide by $\mathbf{\Pi}_{\mathsf{P},n}$, $\mathbf{\Pi}_{\mathsf{U}}$ is an arbitrary permutation matrix that change the order of users. In other words, the beamforming policy is not affected by changing the orders of users, waveguides and
	the pinching antennas in each waveguide.
	
	
	\section{GPASS: Proposed GNN for Learning Beamforming Policy}
	In this section, we design the GPASS architecture to learn the beamforming policy that can leverage the domain knowledge of the policy. 
	
	Intuitively, the input and output of GPASS are respectively the known parameters and decisions of the policy, i.e., $\bm\Phi$ and $\{\mathbf{x}_1^p, \bm\Delta^p,\mathbf{W}\}$. However, simultaneously outputting $\mathbf{x}_1^p, \bm\Delta^p$ and $\mathbf{W}$ may lead to inferior performance because the DNN may not be able to well-capture the coupled relationship between the pinching beamforming and the transmit beamforming matrix. Noticing that once $\mathbf{x}_1^p$ and $\bm\Delta^p$ is known, $\mathbf{G}$ and $\mathbf{H}=[\mathbf{h}_1,\cdots,\mathbf{h}_K]$ are known. Then, problem \textsf{P} reduces to a classical SE-maximal transmit beamforming problem whose solution is relatively easier to learn by leveraging the optimal solution structure in \cite{Precod_Opt_Structure}. Hence, we consider learning the policy in a staged manner, i.e., firstly learning the pinching beamforming with a sub-GNN and then learning the transmit beamforming matrix with another sub-GNN.
	
	\subsection{Learning the Pinching Beamforming}\label{sec:learn-pinching}
	
	We denote $\mathbf{a}_{m_n}\triangleq[x_{1,n}^p, \Delta_{m,n}^p]$, and $\mathbf{A}\triangleq[\mathbf{a}_{1_1}, \cdots,\mathbf{a}_{M_N}]$. The mapping from the user position vector $\bm\Phi$ to the optimal value of $\mathbf{A}$ is denoted as $\mathbf{A}^{\star}=\tilde{F}_1(\mathbf{\Phi})$.
	
	By finding the permutations that do not affect the objective function and constraints in problem \textsf{P}, it can be proved that the mapping $\tilde{F}_1(\cdot)$ satisfies the following permutation property, $\mathbf{\Omega}_{\sf A}^T\mathbf{A}^{\star}=\tilde{F}_1(\mathbf{\Pi}_{\sf U}^T\mathbf{\Phi})$, where $\mathbf{\Omega}_{\sf A}=\mathbf{\Pi}_{\sf A}\otimes\mathsf{diag}(\mathbf{I},\cdots,\mathbf{I})$. $\mathbf{\Omega}_{\sf A}$ is different from $\mathbf{\Omega}_{\sf P}$ that the order of pinching antennas on each waveguide is not changed with $\mathbf{\Omega}_{\sf A}$. The permutation property indicates that the mapping is not affected by changing the orders of waveguides and users, but is affected by changing the order of pinching antennas on each waveguide. This is because the output of the mapping includes the distances between every two adjacent pinching antennas, such that the mapping depends on the order of pinching antennas.
	
	Intuitively, the mapping can be learned over a graph with waveguides and users being two types of vertices, and $\mathbf{A}_n\triangleq[\mathbf{a}_{1_n},\cdots,\mathbf{a}_{M_n}]\in{\mathbb R}^{2\times M}$ is the action of the $n$-th waveguide. However, by doing so, the dimension of action depends on $M$. Hence, the GNN that learns over the graph cannot be adapted to different numbers of $M$. Moreover, the high dimension of $\mathbf{A}_n$ when $M$ is large may incur high training complexity. 
	
	To resolve this issue, we add the index of each pinching antenna into the input of mapping $\tilde{F}_1(\cdot)$ to indicate the order of pinching antennas, which is denoted as $\mathbf{s}=[\mathbf{s}_1^T,\cdots,\mathbf{s}_N^T]^T$, where $\mathbf{s}_n=[s_{1_n},\cdots,s_{M_n}]=[1,\cdots,M]^T$. The mapping now becomes $\mathbf{A}^{\star}=\hat{F}_1(\mathbf{\Phi}, \mathbf{s})$.
	By inputting the indices of pinching antennas, the order of pinching antennas is changeable with the order of indices, indicating that
	the mapping satisfies the following permutation property, 
	\begin{equation}\label{eq:g1-upd}
		\mathbf{\Omega}_{\sf P}^T\mathbf{A}^{\star}=\hat{F}_1(\mathbf{\Pi}_{\sf U}^T\mathbf{\Phi}, \mathbf{\Omega}_{\sf P}^T\mathbf{s}),
	\end{equation}
	i.e., the mapping is not affected by changing the orders of waveguides, pinching antennas on each waveguide, and users.
	
	According to \cite{LSJ_MultiDim_GNN_2022}, the mapping can be learned over a graph with two types of vertices, pinching antenna vertices (PAs) and user vertices (UEs), and the edges are the links from the pinching antennas to the users. The feature of the $k$-th UE is $\bm\psi_k$, and the feature of the $m$-th PA on the $n$-th waveguide (called the $m_n$-th PA) is $s_{m_n}$. The action of the $m_n$-th pinching antenna vertex is $\mathbf{a}_{m_n}$. There are no features and actions on the edges.
	
	We can see that the features and actions are defined on different types of vertices. To avoid the information loss issue such that different features are mapped to the same action, i.e., the features cannot be distinguished, a GNN with $L$ layers should update the representations of edges instead of vertices in each layer with an \emph{update equation} \cite{LSJ_MultiDim_GNN_2022}. Such a GNN is called an \emph{edge-update GNN}. Denote the updated representation of the edge connecting the $k$-th UE and the $m_n$-th PA (called edge $(m_n, k)$) in the $(\ell+1)$-th layer as $\mathbf{d}_{m_n k}^{(\ell+1)}$.
	
	Since the representations of edges are updated in each layer while the features and actions are defined on vertices, the features need to be transformed into the representations of edges in the first layer. Specifically, the representation of each edge is set as the vector of features of the vertices it connected to, i.e., $\mathbf{d}_{m_n k}^{(1)}=[\bm\psi_k, s_{m_n}]$. Moreover, in the last layer, the representations of the edges are transformed into the actions of vertices. Specifically, the action of each PA is obtained by first averaging the representations of edges connected to it, and then passing through a designed activation function to satisfy the constraint in \eqref{eq:opt-constraint-3}, i.e.,
	\begin{equation}\label{eq:act}
		\Delta_{m,n}^p = \max\Big(\frac{1}{K}\sum_{k=1}^K \mathbf{d}_{m_n k}^{(L)}, 0\Big) + \Delta_{\min}.
	\end{equation} 
	
	The updated representations of all the edges in the $(\ell+1)$-th layer constitute a $MN\times K$ matrix $\mathbf{D}^{(\ell+1)}$. Denote the input-output relationship of the update equation in the $\ell$-th layer as $\mathbf{D}^{(\ell+1)}=G(\mathbf{D}^{(\ell)})$. It is not hard to prove that if the following two-dimensional (2D)-PE property is satisfied, i.e., $\mathbf{\Omega}_{\mathsf{P}}^T\mathbf{D}^{(\ell+1)}\mathbf{\Pi}_{\mathsf{UE}}=G(\mathbf{\Omega}_{\mathsf{P}}^T\mathbf{D}^{(\ell)}\mathbf{\Pi}_{\mathsf{UE}})$, then the input-output relationship of the GNN can satisfy the permutation property in \eqref{eq:g1-upd}. 
	
	The update equation can be designed in different forms, and all of them can satisfy the 2D-PE property. Nonetheless, not all of them can be well-generalized to different problem sizes (say the number of users) and trained with low complexity, unless the update equation is judiciously designed. Inspired by the findings of learning beamforming in interference networks in \cite{GJ-RGNN}, as there exists inter-user interference in the PASS, the inter-user interference should be reflected in the update equation of the sub-GNN. To this end, we can design the sub-GNN architecture as follows.
	The representations of all the edges that are connected to the same UE (say the $k$-th UE) are updated together, which are expressed as a vector as $\mathbf{d}_k^{(\ell+1)}=[\mathbf{d}_{1_1 k}^{(\ell+1)T},\cdots,\mathbf{d}_{M_N k}^{(\ell+1)T}]^T$. To update it, the information of representations of edges connected to other UEs are firstly extracted with a \emph{processor} and then aggregated with a \emph{pooling function}. Afterwards, the extracted information is combined with $\mathbf{d}_k^{(\ell)}$ with a \emph{combiner}. The update equation can be expressed as follows,
	\begin{equation}\label{eq:upd-equ}
		\mathbf{d}_k^{(\ell+1)} = \textstyle f\big(\mathbf{d}_k^{(\ell)}, \sum_{j=1,j\neq k}^K q(\mathbf{d}_k^{(\ell)},\mathbf{d}_j^{(\ell)})\big),
	\end{equation}
	where $f(\cdot)$ and $q(\cdot)$ are respectively the combiner and processor that are parameterized functions, and the pooling function is summation $\sum(\cdot)$. 
	In \eqref{eq:upd-equ}, the processor is a function of both $\mathbf{d}_k^{(\ell)}$ and $\mathbf{d}_j^{(\ell)}$, i.e., the representations of edges connected to the $k$-th and the $j$-th user vertices, such that the inter-user interference is reflected.
	
	It is not hard to prove that the input-output relationship of \eqref{eq:upd-equ} is equivariant to arbitrary permutations of UEs, i.e., $\mathbf{D}^{(\ell+1)}\mathbf{\Pi}_{\mathsf{UE}}=G(\mathbf{D}^{(\ell)}\mathbf{\Pi}_{\mathsf{UE}})$ is satisfied. This is because $f(\cdot)$ and $q(\cdot)$ are the same for all the users, and the pooling function $\sum(\cdot)$ satisfies the commutative law. To further guarantee the property to the nested permutations of PAs (which corresponds to the permutations of elements in $\mathbf{d}_k^{(\ell)}$), $\mathbf{y}=f(\mathbf{z})$ and $\mathbf{y}=q(\mathbf{z})$ can be further designed as functions satisfying the nested permutation equivariance property, i.e., $\mathbf\Omega_{\sf P}^T \mathbf{y}=f(\mathbf\Omega_{\sf P}^T \mathbf{z})$ and $\mathbf\Omega_{\sf P}^T\mathbf{y}=q(\mathbf\Omega_{\sf P}^T \mathbf{z})$. To this end, we resort to the following proposition.
	
	\begin{proposition}
		For the mapping $f(\cdot)$ that maps a vector $\mathbf{z}=[\mathbf{z}_1^T,\cdots,\mathbf{z}_N^T]^T$ to another vector $\mathbf{y}=[\mathbf{y}_1^T,\cdots,\mathbf{y}_N^T]^T$, where $\mathbf{z}_n=[z_{1_n},\cdots,z_{{N_p}_n}]^T$, $\mathbf{y}_n=[y_{1_n},\cdots,y_{{N_p}_n}]^T$, $f(\cdot)$ satisfies $\mathbf\Omega_{\sf P}^T \mathbf{y}=f(\mathbf\Omega_{\sf P}^T \mathbf{z})$ if it is with the following form,
		\begin{equation}\label{eq:npe-func}
			y_{m_n} = f_f\big(z_{m_n}, \textstyle\sum_{i=1,i\neq m}^{M} q_{f,1}(z_{i_n}), \textstyle\sum_{j=1,j\neq n}^{N}\sum_{i=1}^{M} q_{f,2}(z_{i_j})\big).
		\end{equation}
		\begin{IEEEproof}
			Due to limited space, the proof is not provided.
		\end{IEEEproof}
	\end{proposition}
	
	In \eqref{eq:npe-func}, $f_f(\cdot), q_{f,1}(\cdot)$ and $q_{f,2}(\cdot)$ can be designed as fully-connected neural networks (FNNs). $q(\cdot)$ can be designed in the same way, i.e., $y_{m_n} = f_q\big(z_{m_n}, \textstyle\sum_{i=1,i\neq m}^{N_p} q_{q,1}(z_{i_n}), \textstyle\sum_{j=1,j\neq n}^{N}\sum_{i=1}^{N_p} q_{q,2}(z_{i_j})\big)$, where $ f_q(\cdot),q_{q,1}(\cdot) $ and $q_{q,2}(\cdot)$ are designed as FNNs. 
	
	We can see that the designed update equation of the GNN is with two recursions. In the first recursion, the update equation of GNN is designed as \eqref{eq:upd-equ} that satisfies the equivariance to the permutations of UEs. In the second recursion, $f(\cdot)$ and $q(\cdot)$ in the update equation are further designed as the form in \eqref{eq:npe-func} to satisfy the nested permutations of PAs.
	
	After obtaining $\Delta_{m,n}^p$ with \eqref{eq:act} in the output layer of the GNN, the $x$-axis position of each pinching antenna can be obtained as $x_{m,n}^p=x_{1,n}^p+\sum_{i=2}^m \Delta_{m,n}^p$. Then, we can obtain $\bm\psi_{m,n}^p=(x_{m,n}^p,y_{n}^p,d)$. Afterwards, $\mathbf{G}$ and $\mathbf{h}_k$ can be obtained with \eqref{eq:pinching-bf} and \eqref{eq:channel-vec}, respectively.
	
	
	We refer to the designed sub-GNN for learning the pinching beamforming (PBF) as PBF-sub-GNN. 
	
	\subsection{Learning the Transmit Beamforming Matrix}\label{sec:learn-bb}
	After $\mathbf{G}$ and $\mathbf{h}_k$ are obtained, problem \textsf{P} reduces to a problem of optimizing transmit beamforming to maximize SE, i.e., 
	\begin{subequations}
		\begin{align}
			\mathsf{Sub-P}: \max_{\mathbf{W}}~ &\sum_{k=1}^K \log_2\Bigg(1+\frac{|\tilde{\mathbf{h}}_k^H\mathbf{w}_k|^2}{\sum_{j=1}^K |\tilde{\mathbf{h}}_k^H\mathbf{w}_j|^2+\sigma_0^2}\Bigg) \label{eq:subopt-objective}\\ 
			{\rm s.t.} ~& \|\mathbf{W}\|^2 \leq P_{\max}, \label{eq:subopt-constraint-1}
		\end{align}
	\end{subequations}
	where $\tilde{\mathbf{h}}_k=\mathbf{h}_k\mathbf{G}$.
	
	It was proved in \cite{Precod_Opt_Structure} that the optimal solution of problem \textsf{Sub-P} is with the following structure, 
	\begin{equation}\label{eq:opt-structure}
		\mathbf{W}^{\star} = \tilde{\mathbf{H}}(\mathbf{\Lambda}\tilde{\mathbf{H}}^{H}\tilde{\mathbf{H}}+\sigma_0^2\mathbf{I}_K)^{-1} \mathbf{P}^{\frac{1}{2}},
	\end{equation}
	where $\tilde{\mathbf{H}}=[\tilde{\mathbf{h}}_1,\cdots,\tilde{\mathbf{h}}_K]$ is the \emph{equivalent channel matrix}, $\mathbf{\Lambda}$ and $\mathbf{P}$ are diagonal matrices with diagonal positions being the uplink and downlink powers allocated to the $K$ users, which are respectively denoted as $\bm\lambda=[\lambda_1,\cdots,\lambda_K]^T$ and $\mathbf{p}=[p_1,\cdots,p_K]^T$.
	
	To simplify the mappings that need to be learned by a GNN, we resort to the solution structure in \eqref{eq:opt-structure}. Specifically, the GNN only learns the uplink and downlink power allocation. Then, the transmit beamforming matrix can be recovered by the solution structure. 
	
	Denote the mapping from the equivalent channel matrix to the power allocation vectors as $\{\mathbf{p}, \bm\lambda\}=F_2(\tilde{\mathbf{H}})$. By finding permutations that do not affect the objective function and constraints of problem \textsf{Sub-P}, it is not hard to prove that the mapping satisfies the following permutation property,
	\begin{equation}\label{eq:pe}
		\{\mathbf{\Pi}_{\sf U}^T\mathbf{p}, \mathbf{\Pi}_{\sf U}^T\bm\lambda\}=F_2(\mathbf{\Pi}_{\sf A}^T\tilde{\mathbf{H}}\mathbf{\Pi}_{\sf U}),
	\end{equation} 
	i.e., the mapping is not affected by changing the orders of waveguides and users. 
	
	The mapping can be learned over the graph with two types of vertices, waveguide vertices (referred to as WGs) and UEs. The edges are the links from the waveguides to the users. The feature of each edge (say the edge from the $n$-th WG to the $k$-th UE, denoted as edge $(n,k)$) is $\tilde{h}_{nk}$, which is the element on the $n$-th row and the $k$-th column of $\tilde{\mathbf{H}}$. There are no features on the vertices. The action of each UE (say the $k$-th UE) is $[p_k, \lambda_k]$, and there are no actions on the WGs and UEs. 
	
	Since the features are defined on edges, an edge-update sub-GNN should be used to learn the mapping from the features to actions to avoid information loss \cite{LSJ_MultiDim_GNN_2022}. As to be validated in section \ref{sec:simulation}, a vanilla edge-update GNN is powerful enough to learn the mapping. The update equation of this GNN can also be written in the form of two recursions as the PBF-sub-GNN.
	Specifically, 
	\begin{itemize}
		\item \emph{First recursion}: The update equation of the GNN can be expressed as,
		\begin{equation}
			\mathbf{d}_k^{(\ell+1)} = \textstyle f\big(\mathbf{d}_k^{(\ell)}, \sum_{j=1,j\neq k}^K q(\mathbf{d}_j^{(\ell)})\big). \notag
		\end{equation}
		\item \emph{Second recursion}: $\mathbf{y}=f(\mathbf{z})$ can be expressed as,
		\begin{equation}\label{eq:pe-func1}
			y_n = \textstyle f_f\big(z_n, \sum_{i=1,i\neq n}^N q_f(z_i)\big), 
		\end{equation}
		where $\mathbf{z}\triangleq [\mathbf{d}_k^{(\ell)}, \sum_{j=1,j\neq k}^K q(\mathbf{d}_j^{(\ell)})], \mathbf{y}\triangleq \mathbf{d}_k^{(\ell+1)}$. $q_f(\cdot)$ is a  parameterized linear function, $f_q(\cdot)$ is a linear function cascaded by an activation function.\\
		$\mathbf{y}=q(\mathbf{z})$ can be expressed as,
		\begin{equation}\label{eq:pe-func2}
			y_n = \textstyle f_q\big(z_n, \sum_{i=1,i\neq n}^N q_q(z_i)\big), 
		\end{equation}
		where $\mathbf{z}\triangleq \mathbf{d}_j^{(\ell)}, \mathbf{y}\triangleq q(\mathbf{d}_j^{(\ell)})$, $f_q(\cdot)$ and $q_q(\cdot)$ are parameterized linear functions.
	\end{itemize}
	
	In this GNN, the processor is only a function of $\mathbf{d}_j^{(\ell)}$, which is different from the PBF-sub-GNN. Moreover, in the second recursion, $f(\cdot)$ in \eqref{eq:pe-func1} and $q(\cdot)$ in \eqref{eq:pe-func2} are functions satisfying $\mathbf{\Pi}_{\sf A}^T \mathbf{y}= f(\mathbf{\Pi}_{\sf A}^T\mathbf{z})$ and $\mathbf{\Pi}_{\sf A}^T \mathbf{y}= q(\mathbf{\Pi}_{\sf A}^T\mathbf{z})$ instead of functions satisfying nested permutation equivariance property as in \eqref{eq:npe-func}.
	
	We refer to the sub-GNN that learns the power allocation and then recovers the transmit beamforming (TBF) matrix with the solution structure in \eqref{eq:opt-structure} as TBF-sub-GNN.
	
	In the output layer of the TBF-sub-GNN, the outputted beamforming matrix $\mathbf{W}$ is passed through an activation function in the following to satisfy the constraint in \eqref{eq:opt-constraint-2},
	\begin{equation}
		\mathbf{W}' = \frac{\mathbf{W}}{\|\mathbf{W}\|}\cdot P_{\max}.
	\end{equation}

\begin{figure*}[!htb]
		\centering
		\includegraphics[width=.9\linewidth]{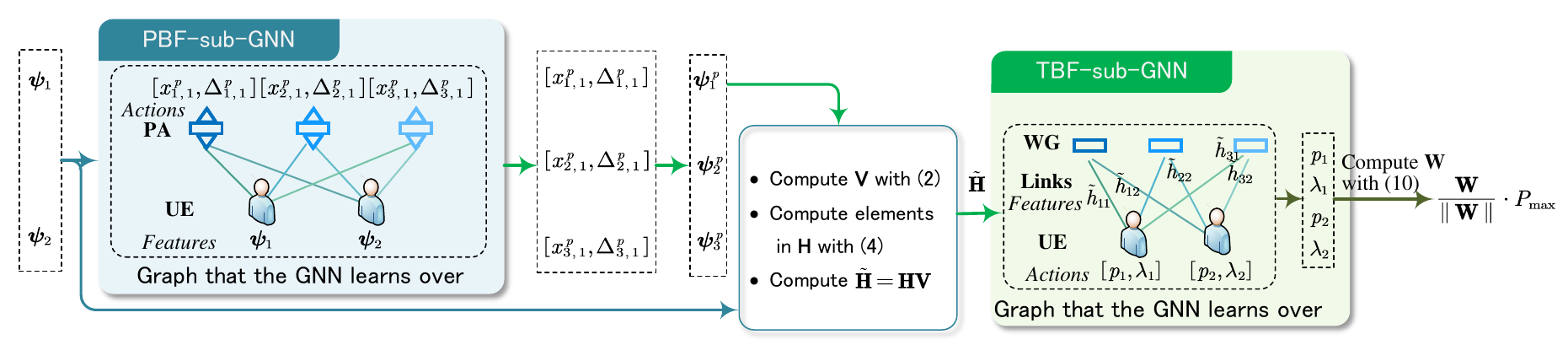}
		\caption{Illustration of GPASS architecture, $N=3, M=1, K=2$.}
		\label{fig:gnn}
		\vspace{-5mm}
	\end{figure*}

	The overall architecture of the GNN is illustrated in Fig. \ref{fig:gnn}. The GNN can be trained in an unsupervised manner, where the loss function is set as the negative SE averaged over all the training samples. 

	\section{Numerical Results}\label{sec:simulation}
	Consider a PASS with $N$ waveguides, and there are $M$ pinching antennas on each waveguide. $K$ users are uniformly distributed in a $10\times 10$ m$^2$ squared region, $d=3$ m, $f_c=28$ GHz, $n_{\rm eff}=1.4$, $\Delta_{\min}=\lambda_g$. Without loss of generality, we set $N=K$.
	
	In the simulations, we generate 10000 samples to train the GPASS. Since it is trained with unsupervised learning, each training sample only contains the input of the GPASS, i.e., the positions of users, which is generated by following the uniform distribution as shown above. After training the GPASS, it is tested on another 1000 samples. 
	
	The learning performance is measured with the SE achieved by the learned beamforming. The performance of the GPASS is compared with a baseline when $M=1$, where the $x$-axis position of each pinching antenna is set as the $x$-axis position of the closest user, and the transmit beamforming is set as zero-forcing. 
	
\begin{figure}[!htb]
		\centering
		\begin{minipage}[t]{0.75\linewidth}	
			\subfigure[$M=1$]{
				\includegraphics[width=\textwidth]{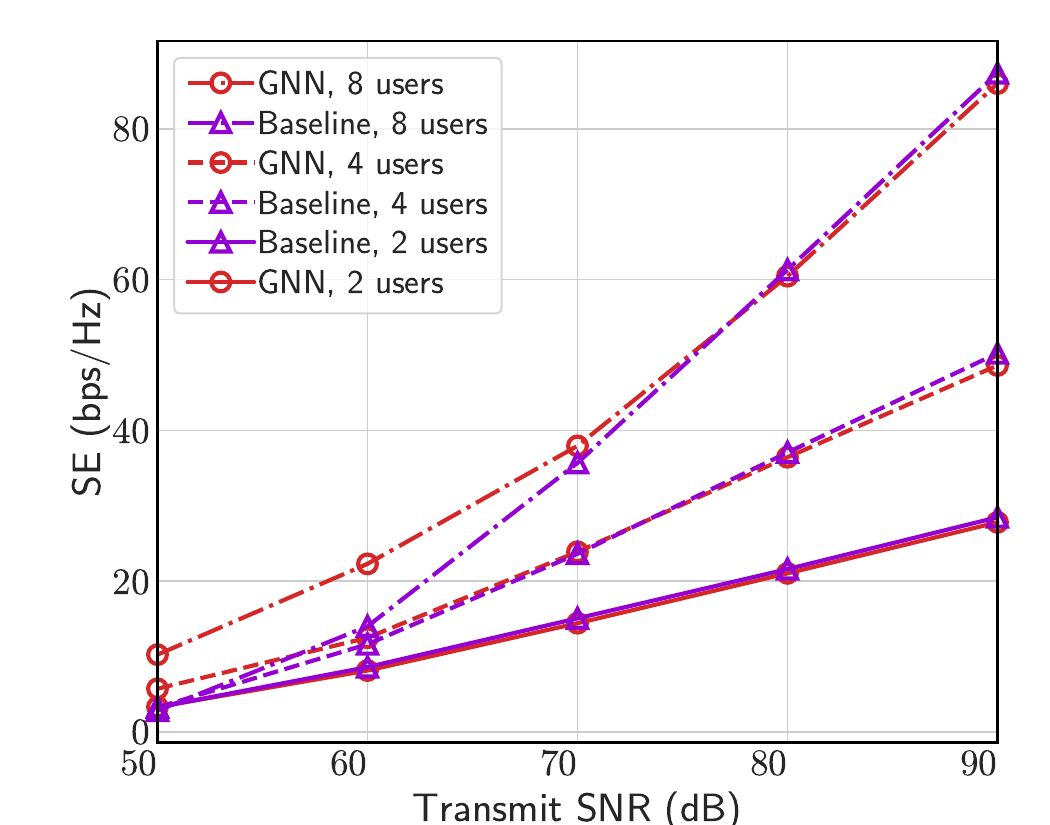}}
		\end{minipage}
		
		\begin{minipage}[t]{0.75\linewidth}	
			\subfigure[$M=3$]{
				\includegraphics[width=\textwidth]{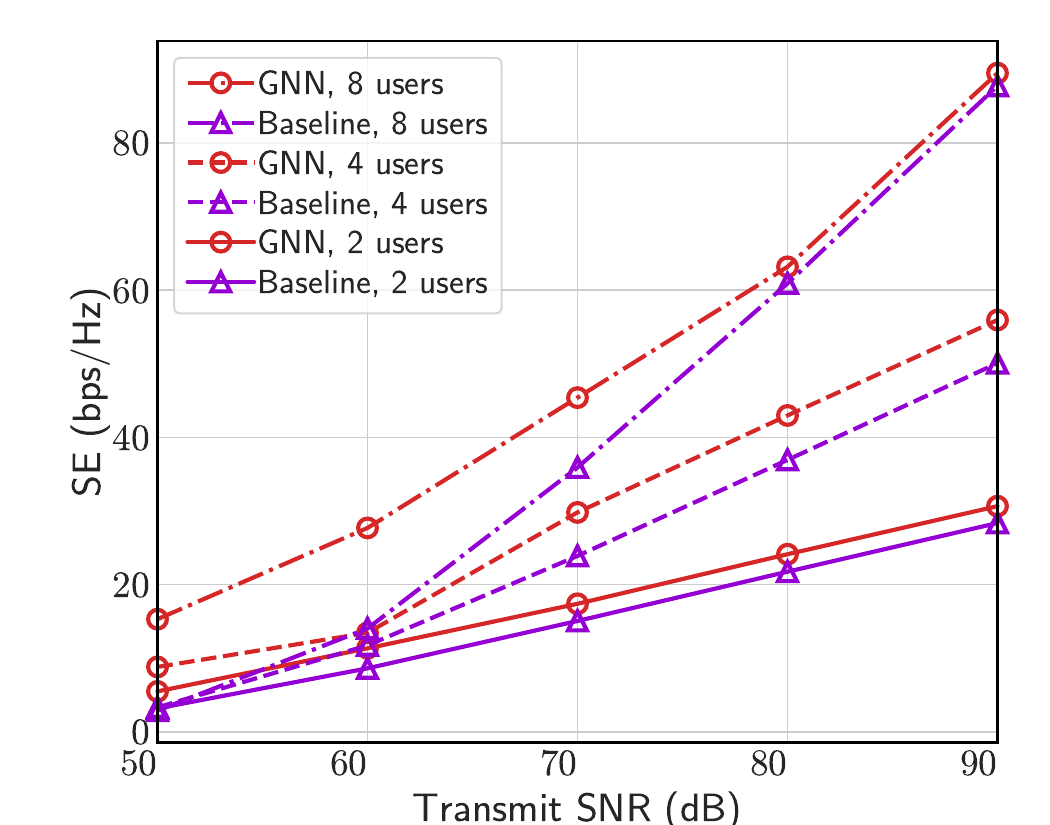}}
		\end{minipage}
		\caption{SE versus SNR.}\label{fig:perf-snr}
	\end{figure}

	In Fig. \ref{fig:perf-snr}, we show the performance of the GPASS in scenarios with different numbers of users and transmit signal-to-noise ratios (SNRs), which is defined as $P_{\max}/\sigma_0^2$. It can be seen that in all the scenarios, the proposed GNN can achieve SE that is close to or higher than the baseline method, and the performance gain is larger when $M=3$.  
	
	We then evaluate the inference time of the proposed GNN. When $M=3, K=N=8$, the inference time averaged over all the test samples is 6 milliseconds, which indicates that the proposed GNN can be implemented in real time.
	
	\section{Conclusions}
	In this letter, we proposed the GPASS architecture for jointly learning the pinching beamforming and transmit beamforming in PASS. The proposed GPASS is with a staged architecture, where the pinching beamforming is first learned. Then, the transmit beamforming matrix is learned by resorting to the optimal solution structure. The permutation properties were incorporated in the GPASS for better learning performance than FNNs. As validated by numerical results, the GPASS can achieve higher SE than the baseline method with a short inference time. The GPASS also has the potential of generalizability to the numbers of pinching antennas, waveguides and users, and joint optimization with channel estimation, which can be investigated in future works.
	
	\bibliography{IEEEabrv,GJ}

\begin{thebibliography}{10}
\providecommand{\url}[1]{#1}
\csname url@samestyle\endcsname
\providecommand{\newblock}{\relax}
\providecommand{\bibinfo}[2]{#2}
\providecommand{\BIBentrySTDinterwordspacing}{\spaceskip=0pt\relax}
\providecommand{\BIBentryALTinterwordstretchfactor}{4}
\providecommand{\BIBentryALTinterwordspacing}{\spaceskip=\fontdimen2\font plus
\BIBentryALTinterwordstretchfactor\fontdimen3\font minus
  \fontdimen4\font\relax}
\providecommand{\BIBforeignlanguage}[2]{{%
\expandafter\ifx\csname l@#1\endcsname\relax
\typeout{** WARNING: IEEEtran.bst: No hyphenation pattern has been}%
\typeout{** loaded for the language `#1'. Using the pattern for}%
\typeout{** the default language instead.}%
\else
\language=\csname l@#1\endcsname
\fi
#2}}
\providecommand{\BIBdecl}{\relax}
\BIBdecl

\bibitem{nearfield-survey}
Y.~Liu, C.~Ouyang, Z.~Wang, J.~Xu, X.~Mu, and A.~L. Swindlehurst, ``Near-field
  communications: A comprehensive survey,'' \emph{IEEE Commun. Surveys Tut.},
  2024, Early access.

\bibitem{ding2024flexible}
Z.~Ding, R.~Schober, and H.~V. Poor, ``Flexible-antenna systems: A
  pinching-antenna perspective,'' \emph{arXiv:2412.02376}, 2024.

\bibitem{ouyang2025array}
C.~Ouyang, Z.~Wang, Y.~Liu, and Z.~Ding, ``Array gain for pinching-antenna
  systems {(PASS)},'' \emph{arXiv preprint arXiv:2501.05657}, 2025.

\bibitem{wang2024antenna}
K.~Wang, Z.~Ding, and R.~Schober, ``Antenna activation for {NOMA} assisted
  pinching-antenna systems,'' \emph{arXiv:2412.13969}, 2024.

\bibitem{liu2025pinching}
Y.~Liu, Z.~Wang, X.~Mu, C.~Ouyang, X.~Xu, and Z.~Ding, ``Pinching antenna
  systems {(PASS)}: Architecture designs, opportunities, and outlook,''
  \emph{arXiv:2501.18409}, 2025.

\bibitem{L2O}
H.~Sun, X.~Chen, Q.~Shi \emph{et~al.}, ``Learning to optimize: training deep
  neural networks for interference management,'' \emph{IEEE Trans. Signal
  Process.}, vol.~66, no.~20, pp. 5438--5453, 2018.

\bibitem{ENGNN}
Y.~Wang, Y.~Li, Q.~Shi, and Y.-C. Wu, ``{ENGNN}: A general edge-update
  empowered {GNN} architecture for radio resource management in wireless
  networks,'' \emph{IEEE Trans. Wireless Commun.}, vol.~23, no.~6, pp.
  5330--5344, June 2024.

\bibitem{GNN_RIS_FL_TWC_2024}
Z.~Wang, Y.~Zhou, Y.~Zou, Q.~An, Y.~Shi, and M.~Bennis, ``A graph neural
  network learning approach to optimize {RIS}-assisted federated learning,''
  \emph{IEEE Trans. Wireless Commun.}, vol.~22, no.~9, pp. 6092--6106, Sep.
  2023.

\bibitem{GJ-RGNN}
J.~Guo and C.~Yang, ``Recursive {GNNs} for learning precoding policies with
  size-generalizability,'' \emph{IEEE Trans. Mach. Learn Commun. Netw.},
  vol.~2, pp. 1558--1579, 2024.

\bibitem{LSJ_MultiDim_GNN_2022}
S.~Liu, J.~Guo, and C.~Yang, ``Multidimensional graph neural networks for
  wireless communications,'' \emph{IEEE Trans. Wireless Commun.}, vol.~23,
  no.~4, pp. 3057--3073, April 2024.

\bibitem{GNN_mUE_RIS_JSAC_2021_Yu}
T.~Jiang, H.~V. Cheng, and W.~Yu, ``Learning to reflect and to beamform for
  intelligent reflecting surface with implicit channel estimation,'' \emph{IEEE
  J. Sel. Areas Commun.}, vol.~39, no.~7, pp. 1931--1945, July 2021.

\bibitem{GNN-PC-CellFree-TWC2024}
S.~Mishra, L.~Salaun, H.~Yang, and C.~S. Chen, ``Graph neural network aided
  power control in partially connected cell-free massive {MIMO},'' \emph{IEEE
  Trans. Wireless Commun.}, vol.~23, no.~9, pp. 12\,412--12\,423, Sep. 2024.

\bibitem{Precod_Opt_Structure}
E.~Bj{\"o}rnson, M.~Bengtsson, and B.~Ottersten, ``Optimal multiuser transmit
  beamforming: A difficult problem with a simple solution structure,''
  \emph{IEEE Signal Process. Mag.}, vol.~31, no.~4, pp. 142--148, July 2014.

\end{thebibliography}
\end{document}